# DEVELOPMENT OF AN E-LEARNING SYSTEM INCORPORATING SEMANTIC WEB

Khurram Naim Shamsi[1], Zafar Iqbal Khan[2]

[1]Lecturer, Dept. of Computer Science, RCC, King Saud University,
Email: kshamsi@ksu.edu.sa

[2]Lecturer, College of Computer Engineering & Sciences, Salman Bin Abdulaziz University,
Email: mzafar.ikhan@gmail.com

**Abstract:** E-Learning is efficient, task relevant and just-in-time learning grown from the learning requirements of the new and dynamically changing world. The term "Semantic Web" covers the steps to create a new WWW architecture that augments the content with formal semantics enabling better possibilities of navigation through the cyberspace and its contents. In this paper, we present the Semantic Web-Based model for our e-learning system taking into account the learning environment at Saudi Arabian universities. The proposed system is mainly based on ontology-based descriptions of content, context and structure of the learning materials. It further provides flexible and personalized access to these learning materials. The framework has been validated by an interview based qualitative method.

**Keywords:** E-learning, Learning Objects, Ontology, RDF, Semantic Web, Software Agents.

## I. INTRODUCTION

The evolution of information society has led to the emergence of new educational technologies and environments. One of the most important requirements to such environments is the rapid (just-in time) access to the relevant knowledge that meets customer's needs as precisely and fully as possible [1]. In recent years the significant progress has been achieved in creation of eLearning systems based on Web technologies. Web-based courses offer clear advantages to learners by allowing very fast, just-in-time, relevant, and at any time or place access to educational resources. However the traditional Web technologies based on a syntactical mark-up of information don't provide the semantic search and navigation within a distributed knowledge environment. This restriction significantly narrows the ability of widespread educational environments to adapt instructional services for a particular customer and preclude from raising the efficiency of learning to a qualitatively new level. To implement such services, the creation of open intelligent educational environments based on Semantic Web is required [2].

## III. SEMANTIC WEB ARCHITECTURE

The term "Semantic Web" encompasses efforts to build a new WWW architecture that supports content

Semantic Web is about building an appropriate infrastructure for intelligent agents to run around the Web performing complex actions for their users [3]. Furthermore, Semantic Web is about explicitly declaring the knowledge embedded in many web-based applications, integrating information in an intelligent way, providing semantic-based access to the Internet, and extracting information from texts [4].

Ultimately, Semantic Web is about how to implement reliable, large-scale interoperation of Web services, to make such services computer interpretable, i.e., to create a Web of machine-understandable and interoperable services that intelligent agents can discover, execute, and compose automatically [5]. Researchers from the World Wide Web Consortium (W3C) already developed new technologies for web friendly data description [6]. Moreover, AI researchers have already developed some useful applications and tools for the Semantic Web [7].

This paper outlines how the Semantic Web can be used as a technology for realizing sophisticated eLearning scenarios. In the following, we will give a brief overview about the Semantic Web and discuss a number of important issues. We mention the layers of the Semantic Web architecture. In the subsequent section, we introduce the framework of our Semantic Web-based e-learning system. We continue with a description of an ontology based approach for eLearning. After a discussion of the model, concluding remarks summarize the importance of the presented topics and outline some future work.

## II. SEMANTIC WEB

Semantic Web (SW) derives from W3C director Tim Berners-Lee's vision of Web as a universal medium for data, information and knowledge exchange [8]. The word semantic web is a product of Web2.0 (second generation web) which makes the web itself to understand and satisfy the user requests and web agents or machines to use the content of web [9][10].

with formal semantics. It means the content suitable for automated systems to consume contrary to the content intended for human consumption. This enables automated agents to reason about the Web content, and produces an intelligent response to unforeseen situations.

"Expressing meaning" is the most important feature of the Semantic Web. In order to accomplish this goal, several layers of representational structures are needed. They are presented in the Figure 1 [11] below, among which the following layers are the basic ones:

- the XML layer, which represents the structure of data;
- the RDF layer, which represents the meaning of data;
- the Ontology layer, which represents the formal common agreement about meaning of data;
- the Logic layer, which enables intelligent reasoning with meaningful data.

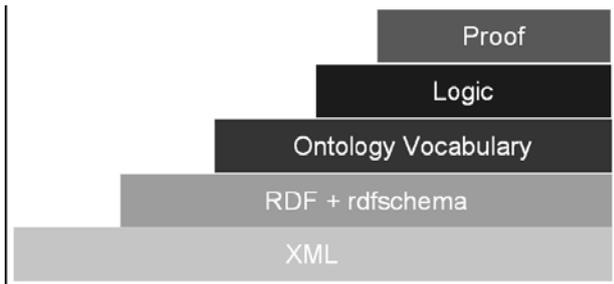

*Figure 1: Layers of the semantic web architecture*

## IV. SEMANTIC WEB MINING AND E-LEARNING

The term Semantic Web Mining is described well by Stumme and etc. all. [12] as "Semantic Web Mining aims at combining the two areas Semantic Web and Web Mining. This vision follows our observation that trends converge in both areas: Increasing number of researchers work on improving the results of Web Mining by exploiting semantic structures in the Web, and make use of Web Mining techniques that can be used for mining Semantic Web itself. The wording Semantic Web Mining emphasizes this spectrum of possible interaction between both research areas: It can be read both as Semantic (Web Mining) and as (Semantic Web) Mining."

Pointing at the definitions given, it's possible to use web logs for any course available on any course management system or e-learning portal for investigation of semantic information. In a case study on Moodle, case studies for applications of data mining techniques are given by Romero, Ventura, & Garcia [13]. In these studies, the possible techniques for data retrieval and management, educator has to run third party programs manually for information retrieval, for educator are explained briefly. For a semantic and real time system, web services and web agents were announced to be useful. Also, trustworthiness of the data is very important since it can lead the algorithms or mining techniques in wrong

*A. Knowledge Base*

It is a repository where ontologies, metadata, inference rules, educational resources and course or inadequate results. At this point, we can assume that the data we get from student's answers is reliable or we can also run data mining algorithms to fetch conflictions on answers to filter them somehow.

## V. SEMANTIC BASED CONCEPTUAL ELEARNING PLATFORM ARCHITECTURE

Figure 2 depicts a conceptual Semantic eLearning framework that provides high-level services to people looking for appropriate online information. The basic levels of this framework are the human levels composed by the Students and the Instructors, the access level that grant access to students and instructors into the system, the interface level that provide various facilities, the service level that handles the background processes, and the knowledge base level comprising of various repositories. In the base of this conceptual architecture, we have depicted the key elements of a semantic eLearning platform.

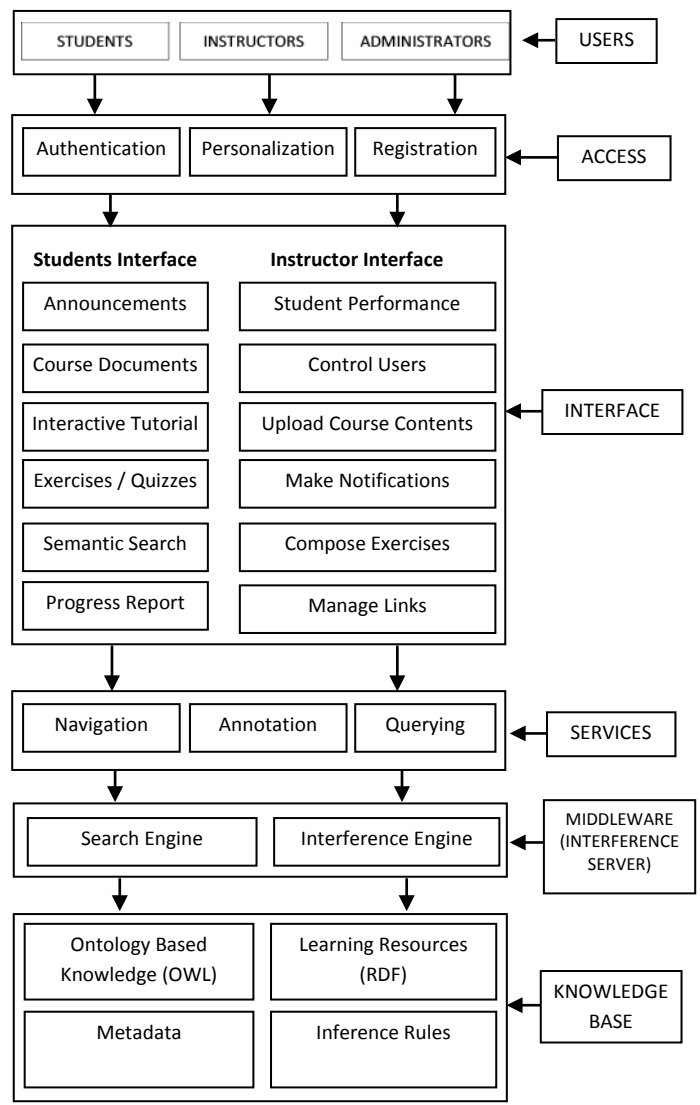

*Figure 2: Conceptual Architecture for Semantic eLearning*

descriptions, user profiles, etc. are stored. The metadata may be placed within the document itself or in some external metadata repository [14]. In the Figure 2 the metadata are stored externally in the knowledge base because it is easier to scan a separate Meta description stored in a database and it takes less space to store it. The second advantage is that the point of view may vary according to different authors who reuse the same learning material. It means that it is possible to have different descriptions of the learning material according to the different contexts.

*B. Search Engine*

It provides an API with methods for querying the knowledge base. RDQL (RDF Data Query Language) can be used as ontology query language.

*C. Access layer*

It acts as a security layer between the users and the system. It provides an interface and grant access into the system.

*D. Interface layer*

It provides an integrated interface through which students as well as Instructors / Administrators of academic institutions can access, upload or modify the data with particular authority. The users can manage different services and the instructors and administrators can control data and users.

*E. Ontology Based Knowledge*

In an eLearning environment the situation can easily arise that different instructors use different terminologies, in which case the combination of learning materials becomes difficult. The retrieval problem is additionally compounded by the fact that typically instructors and students have very different backgrounds and levels of knowledge. Therefore, some mechanism for establishing a shared-understanding is needed. Ontologies are a powerful mechanism for achieving this task. In fact, ontology constrains the set of possible mapping between symbols and their full meanings [15].

*F. Annotations*

Annotation is the activity of annotating text documents written in plain ASCII or HTML with a set of tags representing the names of slots of the selected class in ontology. Semantic annotation of WEB resources requires the ability to convert syntactical information into semantic descriptors referred to a conceptual domain model [16]. However the selection of relevant annotations can be very expensive in terms of complexity since it requires understanding semantic relationships between concepts used to describe web resources; it requires the ability to cope with different levels in document descriptions and it should produce semantically significant mapping of resources to concepts. The annotation can be manual or automatic. Several approaches have been proposed for automatic semantic annotation, rooted in two main research fields: natural language processing and machine learning. Natural language processing (NLP) is based on the detection of typical human constructs in textual information and tries to map known sentence composition rules to semantic rich descriptions. The other way to provide semantic descriptors currently under investigation is machine learning. Basically machine learning means extracting association rules and behaviours to allow machines to accomplish specific tasks as well as their human counterparts.

*G. Metadata*

Metadata is data about data that helps us to achieve better search results [14]. Each component of the eLearning system can be described with the help of metadata. The metadata level is the first level of a semantic WEB-based application [17]. This metadata can be attached to each software component of the eLearning system in order to store several important characteristics (e.g., information regarding uptime, ownership, execution platform, etc.). Also, for each user we can retain the information about his/her status. For example, we can store the user role – administrator, database manager, security monitor, regular user. Also, the system can retain personal data (e.g., age, user e-mail address, location, etc.). Instead of hoping that a full text search through a learning resource will find the author's name Henry for example, we can annotate the resource with a metadata description "author is Henry". We can also easily realize that there are two major difficulties in this method. The first difficulty is the technical realization of attaching metadata at a resource and the second difficulty is the standardization of descriptions in order to avoid misunderstanding by using different attributes for the same purpose like "creator is Henry" or "written by Henry". The solution for the first problem is the usage of two possible approaches that have been developed in the context of the World Wide Web, based on the XML and RDF formalisms. The solution for the second problem is the usage of standard vocabularies or schemas for metadata to describe digital resources.

## VI. METHODOLOGY

The present framework is quite simple and general in nature in a way that it can be applied to any higher education system. However, while designing the framework, main focus was on the requirements of higher learning institutions in Saudi Arabia. In order to prove its worthiness for Saudi universities, an interview based qualitative method was utilized. The detailed framework was provided to key figures in the Saudi universities such as deans and other administrative staffs in the deanship of Graduate

Studies, deanship of e-transaction, deanship of e-learning and distance education, etc. and their views were sought. Most of the participants gave their supporting views. Some of them suggested certain improvements that were incorporated after due discussion.

## VII. CONCLUSION AND FUTURE WORK

The main contribution of this paper is our new model for e-learning system, using the Semantic Web technology. The framework includes various services and tools in the context of a semantic portal, such as: registration, uploading course documents, Interactive tutorial, announcements, notifications, and simple semantic search. A metadata-based ontology is introduced for this purpose and added to our model. The OWL language is used to develop our ontologies. In these ontologies, the actual resources and properties specified in the RDF models are defined. The Protégé2000 ontology editor can be used to create the e-learning ontology classes and properties. A list of the technologies required for the implementation of our web-based e-learning system includes PHP Platform, Apache Web Server, MySQL database, and RAP Semantic Web Toolkit.

We believe that there are two primary advantages of our Semantic web-based model. One is that the proposed model, which contains a hierarchical contents structure and semantic relationships between concepts, can provide related useful information for searching and sequencing learning resources in web-based eLearning systems. The other is that it can help a developer or an instructor to develop a learning sequence plan by helping the instructor understand the why and how of the learning process.